\documentclass[a4paper]{PoS}
\usepackage{graphicx}
\usepackage{amsbsy}

\def\cp{$CP$\/}
\def\cpb{$\pmb{CP}$\/}

\def\ra{\!\rightarrow\!}
\def\dbar{\overline{D}{}^{\,0}}

\def\af{{\cal A}^{}_f}
\def\abarf{\overline{\cal A}^{}_f}

\def\simge{\mathrel{%
   \rlap{\raise 0.511ex \hbox{$>$}}{\lower 0.511ex \hbox{$\sim$}}}}
\def\simle{\mathrel{
   \rlap{\raise 0.511ex \hbox{$<$}}{\lower 0.511ex \hbox{$\sim$}}}}

\title{
\vskip-1.0in
\begin{flushright}
\normalsize{\rm
UCHEP--17--05 \\
26 April 2017
}
\end{flushright}
\vskip1.0in
Prospects for time-dependent mixing and 
\cpb-violation measurements at Belle~II
}

\ShortTitle{Time-dependent mixing and \cp\ violation at Belle~II}

\author{\speaker{A. J. Schwartz}\thanks{On behalf of the Belle~II Collaboration}\\
        Physics Department, University of Cincinnati, Cincinnati, Ohio 45221\\
        E-mail: \email{alan.j.schwartz@uc.edu}}

\abstract{
The Belle~II experiment is under construction at the
KEK laboratory in Japan. Belle~II will study $e^+e^-$ collisions
at or near the $\Upsilon(4S)$ resonance with the goal of collecting
50~ab$^{-1}$ of data, which is a large increase over that recorded by 
the Belle and BaBar experiments. This data will provide a large 
sample of charm meson decays. In this report we present the expected 
sensitivity of Belle~II for measuring time-dependent mixing and \cp\ 
violation in the $D^0$-$\dbar$ system. We focus on measurements of 
$D^0\ra K^+\pi^-$, $D^0\ra K^+\pi^-\pi^0$, and
$D^0\ra K^0_S\pi^+\pi^-$ decays.
}

\FullConference{9th International Workshop on the CKM Unitarity Triangle\\
		28  November - 3 December 2016\\
		Tata Institute for Fundamental Research (TIFR), Mumbai, India}

\begin{document}

\section{Introduction}

The Belle experiment at the KEK laboratory in Japan is now
being upgraded to the Belle~II experiment. The goal of Belle~II 
is to collect a data set corresponding to 50 times what Belle obtained. 
The experiment is designed to search for new physics in $B$ and
$D$ meson decays, and in $\tau$ lepton decays. Within this broad physics
program, measuring mixing and \cp\ violation in the neutral charm
system is an important goal. The first evidence for $D^0$-$\dbar$ 
mixing was obtained in 2007 by the Belle~\cite{belle_mixing} and 
BaBar~\cite{babar_mixing} experiments, with subsequent observations
made by LHCb~\cite{lhcb_mixing} and CDF~\cite{cdf_mixing}. However, 
the current precision on the mixing parameters $x= \Delta m/\Gamma$ 
and $y= \Delta\Gamma/(2\Gamma)$, where $\Delta m$ and $\Delta \Gamma$ 
are the differences in masses and decay widths between the two 
mass eigenstates and $\Gamma$ is the mean decay width, is 
insufficient to determine whether $y>x$ or even whether $x\neq 0$~\cite{hflav}.
To establish this, higher statistics measurements are needed.
\cp\ violation in the $D^0$-$\dbar$ system has not yet been observed. 
The predicted rate within the Standard Model (SM) is very
small, and an observation of \cp\ violation by Belle~II 
could indicate new physics.

There are several advantages of performing these measurements at
an $e^+e^-$ experiment such as Belle~II rather than at a hadron machine:
lower backgrounds, higher trigger efficiency, excellent 
$\gamma$ and $\pi^0$ reconstruction (and thus $\eta$, $\eta'$, and 
$\rho^+$ reconstruction), high flavor-tagging efficiency with low 
dilution, and numerous control samples with which to study systematics.
In this report we present Belle~II prospects for measuring the mixing 
parameters $x$ and $y$, and \cp-violating parameters $|q/p|$ 
and ${\rm Arg}(q/p)=\phi$, where $q$ and $p$ are the complex coefficients
relating the $D^0$-$\dbar$ flavor eigenstates with the mass eigenstates.
We focus on three benchmark decay modes: 
$D^0\ra K^+\pi^-$, $D^0\ra K^+\pi^-\pi^0$, and
$D^0\ra K^0_S\pi^+\pi^-$~\cite{charge-conjugates}. In all cases 
the flavor of the neutral $D$ is determined by requiring that it 
originate from a $D^{*+}\ra D^0\pi^+$ or $D^{*-}\ra\dbar\pi^-$ 
decay; the charge of the $\pi^\pm$ identifies the $D^0$ or 
$\dbar$ flavor. All results given are preliminary.

\section{Decay-time resolution}
\label{section:decay-time}

\begin{figure}[htb]
\hbox{
\includegraphics[width=0.37\textwidth,angle=-90]{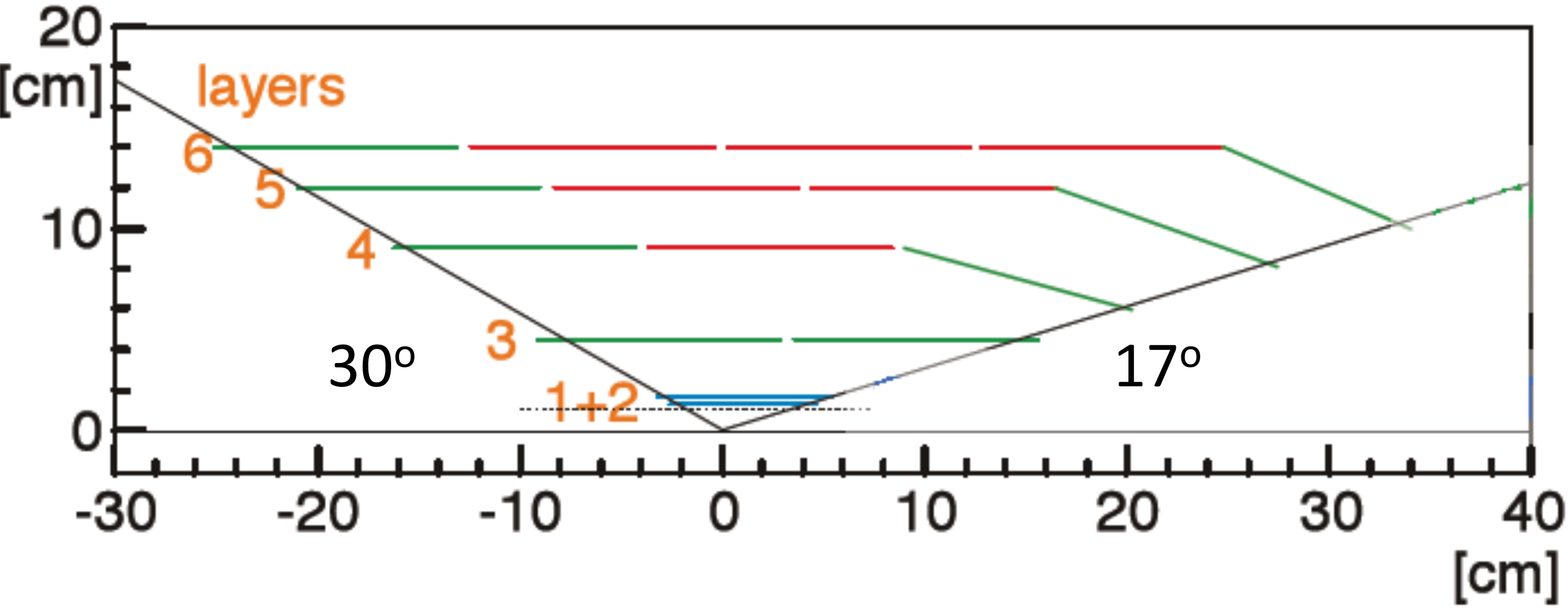}
\hskip0.40in
\includegraphics[width=0.32\textwidth,angle=-90]{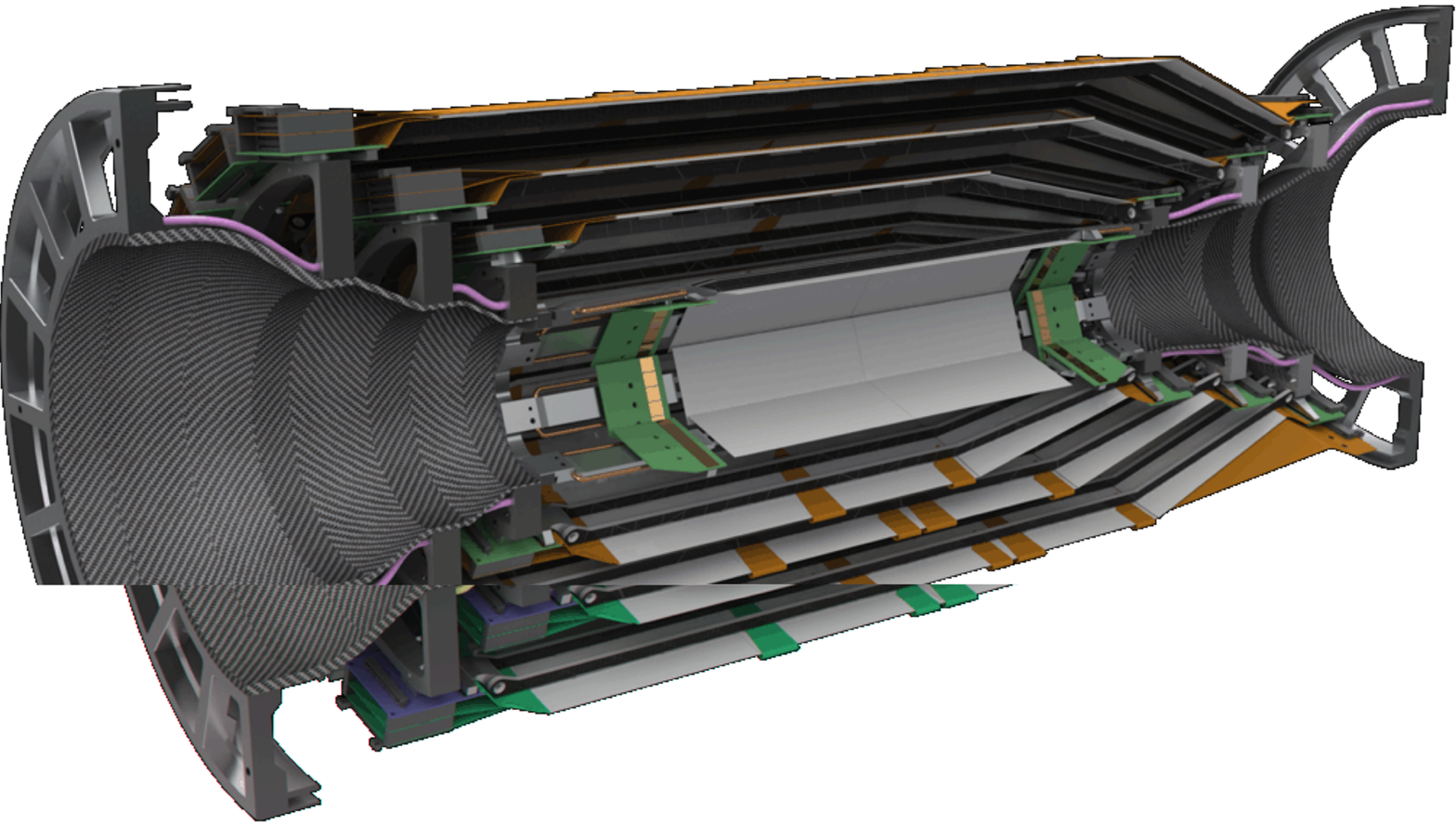}
}
\vskip-0.35in
\caption{
Left: layers of the Belle~II vertex detector, which consists of 
two layers of silicon pixels and four layers of silicon strips.
Right: a cutaway view of the vertex detector.}
\label{fig:belleII_svd}.
\end{figure}

Measurements of time-dependent mixing and \cp\ violation 
depend on measuring decay times precisely. Due to an 
improved vertex detector,  the decay time resolution of 
Belle~II should be superior to that of Belle and BaBar. 
Whereas Belle used a four-layer silicon-strip detector, 
Belle~II will use four layers of silicon strips 
plus two layers of silicon pixels, as shown in Fig.~\ref{fig:belleII_svd}.
Some parameters of the vertex detector are listed in 
Table~\ref{tab:svd_layers}.
The pixel layers will lie only 14~mm and 22~mm from 
the interaction point (IP), and the smallest pixel 
size will be $55\times 50$~$\mu$m$^2$.

\begin{table}[hbt]
\begin{center}
\begin{tabular}{c|cccc}  
Layer &  Type &  Distance from IP & Size or pitch &  Number of pixels \\
      &       &   (mm)            &   ($\mu$m)    &    or strips      \\
\hline
 1 & pixels & 14 & $(55,60)\times 50$ & $3\,072\,000$ \\ 
 2 & pixels & 22 & $(70,85)\times 50$ & $4\,608\,000$ \\
\hline
 3 & strips & 38  &  $50\,(\phi),\,160\,(z)$ &  $768\,(\phi),\,768\,(z)$ \\
 4 & strips & 80  &  $50,75\,(\phi),\,240\,(z)$ &  $768\,(\phi),\,512\,(z)$ \\
 5 & strips & 104 &  $50,75\,(\phi),\,240\,(z)$ &  $768\,(\phi),\,512\,(z)$ \\
 6 & strips & 135 &  $50,75\,(\phi),\,240\,(z)$ &  $768\,(\phi),\,512\,(z)$ \\
\hline
\end{tabular}
\end{center}
\vskip-0.10in
\caption{Layers and segmentation of the Belle~II vertex detector.}
\label{tab:svd_layers}
\end{table}

\begin{figure}[htb]
\hbox{
\includegraphics[width=0.36\textwidth,angle=-90]{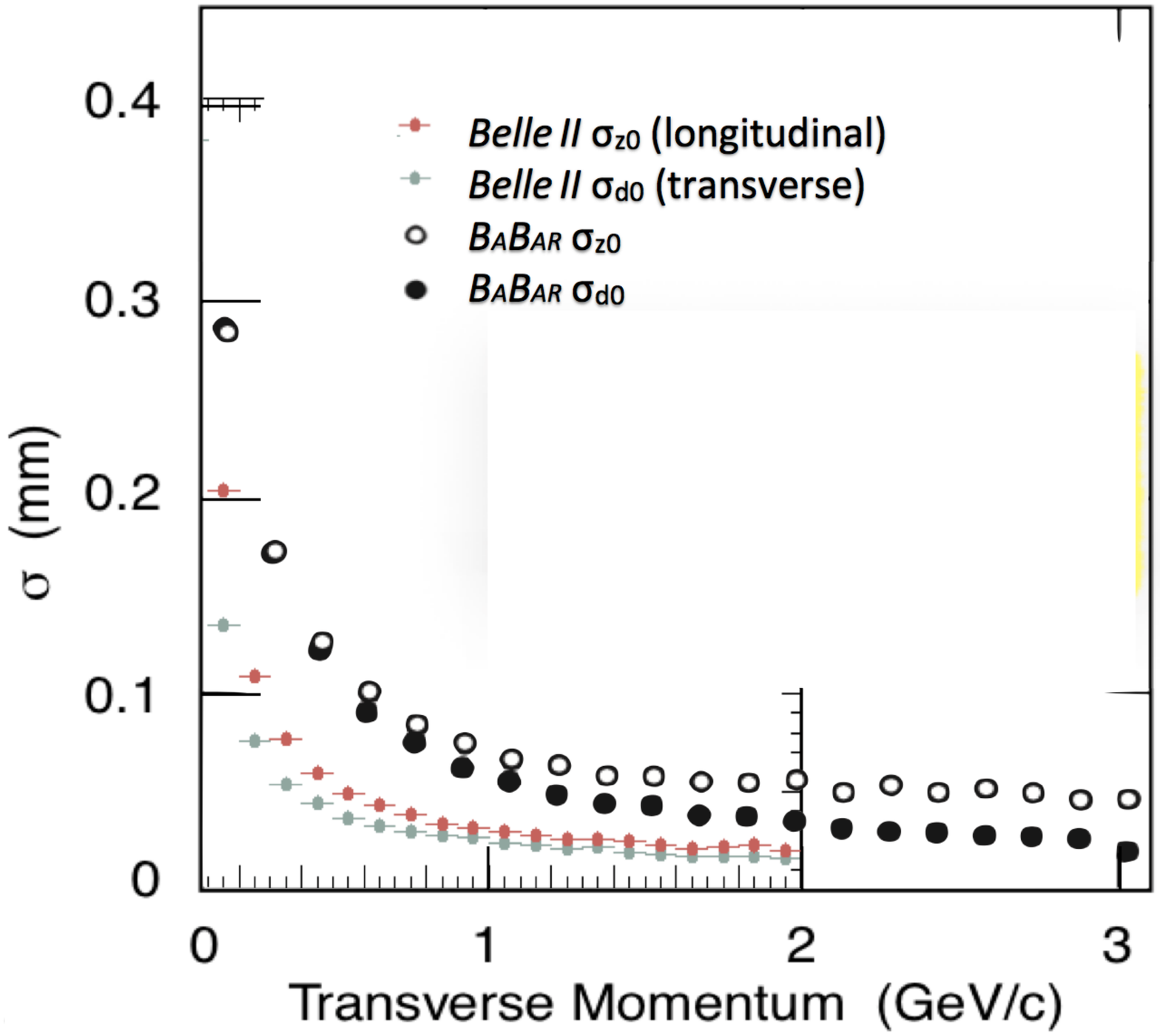}
\hskip0.20in
\includegraphics[width=0.36\textwidth,angle=-90]{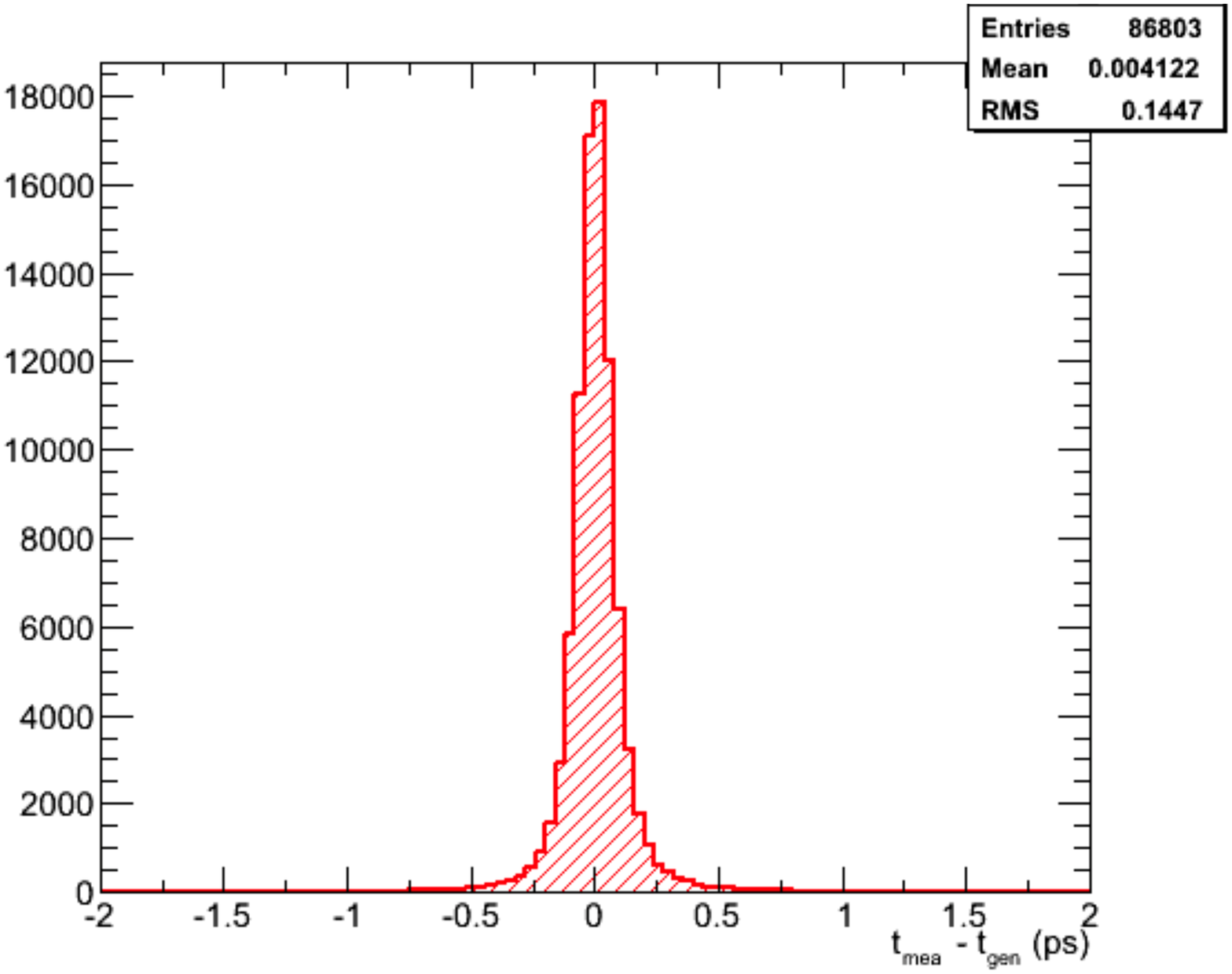}
}
\vskip0.10in
\caption{
Left: Belle~II track impact parameter with respect to the IP as 
a function of track momentum, as obtained from MC simulation.
Right: Belle~II residuals of decay vertex position, from MC
simulation of $D^{*+}\ra D^0\pi^+,\ D^0\ra K^+K^-$ decays. 
The RMS of this distribution is 140~fs.}
\label{fig:resolution}
\end{figure}

Figure~\ref{fig:resolution}~(left)
shows the resolution on the impact parameter of tracks with respect 
to the IP as a function of track momentum, as obtained from Monte 
Carlo (MC) simulation. Also shown are corresponding results from 
BaBar. The figure indicates that the Belle~II resolution will be 
approximately half that of BaBar. 
Figure~\ref{fig:resolution}~(right) plots the residuals of decay 
time for a large sample of MC $D^{*+}\ra D^0\pi^+,\ D^0\ra K^+K^-$ 
decays. The RMS of this distribution is 140~fs, which is almost 
half the decay time resolution of BaBar (270~fs). Similar results
have been obtained for prompt $D^0\ra K^+K^-$ decays, and also for
$D^0\ra\pi^+\pi^-$ decays.

\section{Mixing and \cpb\ violation in $\pmb{D^0\ra K^+\pi^-}$}

To study the sensitivity of Belle~II to parameters 
$x$, $y$, $|q/p|$, and $\phi$ in ``wrong-sign'' $D^0\ra K^+\pi^-$ 
decays, we perform a ``toy'' MC study as follows. 
We generate an ensemble of 1000 MC experiments, with each 
experiment consisting of a sample of $D^0$ decays and a 
separate sample of $\dbar$ decays. The number of events in 
each sample corresponds to 5~ab$^{-1}$, 20~ab$^{-1}$, and 
50~ab$^{-1}$ of data. For the full dataset (50~ab$^{-1}$), 
there are a total of 438600 $D\ra K^\pm\pi^\mp$ decays. 
The decay times are obtained by sampling from the following 
probability density functions (PDFs):
\begin{eqnarray}
\frac{dN}{dt}(D^0) & = &  
e^{-\Gamma t}\left\{
R^{}_D + \left|\frac{q}{p}\right|\sqrt{R^{}_D}(y'\cos\phi - x'\sin\phi)(\Gamma t) +
\left|\frac{q}{p}\right|^2\frac{x'^2+y'^2}{4} (\Gamma t)^2\right\} 
\label{eqn:dkp_decay_time_d0} \\
\frac{dN}{dt}(\dbar) & = &  
e^{-\Gamma t}\left\{
\overline{R}^{}_D + \left|\frac{p}{q}\right|
\sqrt{\overline{R}^{}_D}(y'\cos\phi + x'\sin\phi)(\Gamma t) +
\left|\frac{p}{q}\right|^2\frac{x'^2+y'^2}{4} (\Gamma t)^2\right\}\,,
\label{eqn:dkp_decay_time_d0b}
\end{eqnarray}
where $x' = x\cos\delta + y\sin\delta$,  
$y' = -x\sin\delta + y\cos\delta$, and 
$\delta$ is the strong phase difference between 
$D^0\ra K^-\pi^+$ and $\dbar\ra K^-\pi^+$ amplitudes.
This strong phase difference can be measured at BESIII 
using \cp-tagged $D^{}_{CP}\ra K^-\pi^+$ decays~\cite{besIII_delta}.
The parameter $R^{}_D$ ($\overline{R}^{}_D$) is the ratio
of amplitudes squared
$|{\cal A}(D^0\ra K^+\pi^-)/{\cal A}(D^0\ra K^-\pi^+)|^2$
($|{\cal A}(\dbar\ra K^-\pi^+)/{\cal A}(\dbar\ra K^+\pi^-)|^2$).

After generation, the decay times are smeared by the expected 
decay time resolution of 140~fs, and the resulting $D^0$ and $\dbar$ 
time distributions are simultaneously fitted. Backgrounds are not
yet included in this study; a first look at backgrounds indicates 
that when they are included, the fitted errors increase by 
only a small amount. The PDFs used for the fit are 
the convolution of 
Eqs.~(\ref{eqn:dkp_decay_time_d0}) and (\ref{eqn:dkp_decay_time_d0b})
with a Gaussian resolution function; this results in
products of exponential functions and Error functions.
The preliminary results are listed in Table~\ref{tab:dkp_results}.
The final Belle~II sensitivity for $|q/p|$ is~$<0.1\%$, and that 
for $\phi$ is $6^\circ$. This precision is a significant 
improvement over that which Belle and BaBar achieved.

\begin{table}[hbt]
\begin{center}
\begin{tabular}{l|lll}  
Parameter &  5~ab$^{-1}$ &  20~ab$^{-1}$ &  50~ab$^{-1}$ \\
\hline
$\delta x'$ (\%) & 0.37 & 0.23 & 0.15 \\
$\delta y'$ (\%) & 0.26 & 0.17 & 0.10 \\
$\delta |q/p|$ (\%) & 0.20 & 0.09 & 0.05 \\
$\delta \phi$ ($^\circ$) & 16 & 9.2 & 5.7 \\
\hline
\end{tabular}
\end{center}
\vskip-0.10in
\caption{Preliminary results of a toy MC study of $D^0\ra K^+\pi^-$ 
decays: uncertainty on ``rotated'' mixing parameters $x'$ and $y'$, 
and on \cp-violating parameters $|q/p|$ and $\phi$, for three values 
of integrated luminosity.}
\label{tab:dkp_results}
\end{table}

\section{Mixing in $\pmb{D^0\ra K^+\pi^-\pi^0}$}

We study mixing in $D^0\ra K^+\pi^-\pi^0$ decays using an 
MC simulation based on EVTGEN~\cite{EVTGEN}. For this study 
we generate an ensemble of 10 experiments, 
with each experiment consisting of 225000 $D^0\ra K^+\pi^-\pi^0$ decays 
corresponding to 50~ab$^{-1}$ of Belle~II data. The generated decay 
times are smeared by a resolution of 140~fs, and the decay model 
used to generate and fit the Dalitz 
plot is the isobar model used by BaBar~\cite{babar_dkpp_dalitz}.
Possible \cp\ violation and backgrounds are neglected in this phase
of the study. More details are given in Ref.~\cite{longki_CPC}.

\begin{figure}[htb]
\vskip-1.40in
\includegraphics[width=0.78\textwidth,angle=-90]{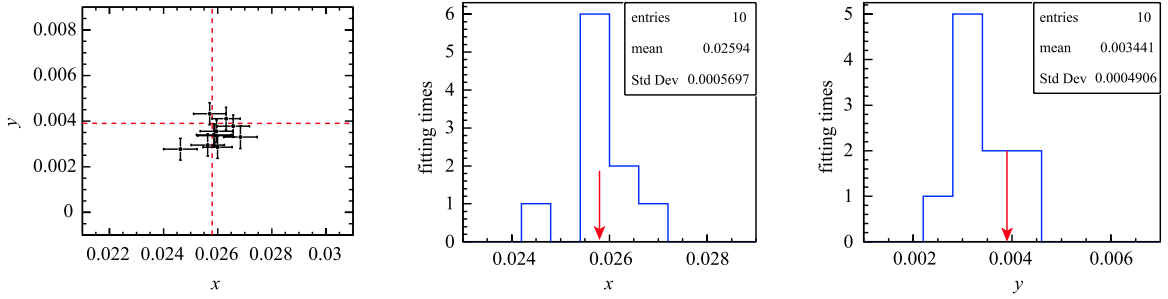}
\vskip-1.60in
\caption{
Left: preliminary results of fitting an ensemble of 10 
experiments, with each experiment corresponding to 50~ab$^{-1}$ 
of data~\cite{longki_CPC}.
Middle and right: projections of the left-most plot. The RMS 
values of these distributions are $\delta x = 0.057\%$ and 
$\delta y = 0.049\%$.}
\label{fig:kpp_dalitz_results}
\end{figure}

\begin{figure}[htb]
\vskip-1.60in
\includegraphics[width=0.78\textwidth,angle=-90]{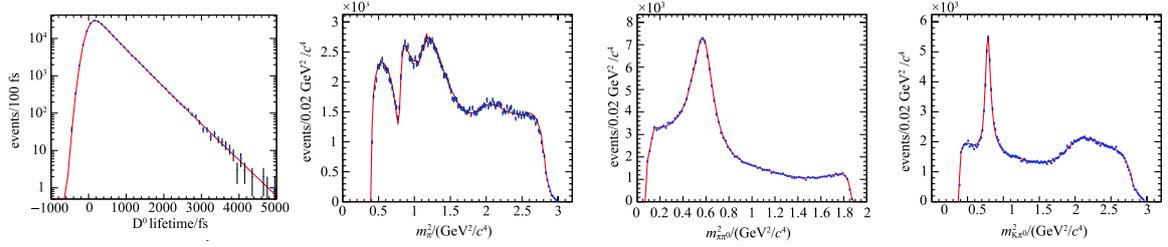}
\vskip-1.74in
\caption{
Projections of the fit to $D^0\ra K^+\pi^-\pi^0$ events
for one typical experiment in the ensemble~\cite{longki_CPC}.}
\label{fig:kpp_dalitz_onefit}
\end{figure}

The mixing parameters are 
$x'' = x\cos\delta^{}_{K\pi\pi} + y\sin\delta^{}_{K\pi\pi}$ and 
$y'' = -x\sin\delta^{}_{K\pi\pi} + y\cos\delta^{}_{K\pi\pi}$, where
$\delta^{}_{K\pi\pi}$ is the strong phase difference between 
$D^0\ra K^-\pi^+\pi^0$ and $\dbar\ra K^-\pi^+\pi^0$ amplitudes
evaluated at $m^{}_{\pi\pi} = M^{}_\rho$. For this study the
strong phase $\delta^{}_{K\pi\pi}$ is fixed to $10^\circ$, 
and the fitted parameters are $x$ and $y$. 
The results are shown in Fig.~\ref{fig:kpp_dalitz_results}. 
The input values are $x = 0.0258$ and $y = 0.0039$, and the 
RMS of the distributions of residuals are $\delta x = 0.057\%$ 
and $\delta y = 0.049\%$. This precision is approximately 
one order of magnitude better than that achieved by 
BaBar~\cite{babar_dkpp_dalitz}. 
The projections of a typical fit (one experiment in the 
ensemble) are shown in Fig.~\ref{fig:kpp_dalitz_onefit}.
In practice, to extract values for $x$ and $y$ requires 
knowledge of the strong phase $\delta^{}_{K\pi\pi}$; this 
can in principle be measured at BESIII using \cp-tagged 
$D^{}_{CP}\ra K^-\pi^+\pi^0$ decays. 

\section{Mixing and \cpb\ violation in $\pmb{D^0\ra K^0_S\pi^+\pi^-}$}

Another method for measuring mixing and \cp\ violation
is by fitting the time-dependent Dalitz plot for 
$D^0\ra K^0_S\pi^+\pi^-$ decays. One calculates the 
observables 
$m^{}_+\equiv (P^0_{K_S} + P^{}_{\pi^+})^2$ and
$m^{}_-\equiv (P^0_{K_S} + P^{}_{\pi^-})^2$ and 
performs an unbinned maximum likelihood fit 
to $m^{}_+$, $m^{}_-$, and the decay time $t$.
To fit for \cp-violating parameters $|q/p|$ 
and $\phi$ (in addition to $x$ and $y$) requires 
dividing the data sample into $D^0$ and $\dbar$ 
decays and fitting the two subsamples simultaneously. 

The time-dependent PDFs are 
\begin{eqnarray}
\frac{dN}{dt}(D^0) & = &  
\frac{e^{-\Gamma t}}{2}
\left\{ \left(|\af|^2 + \left|\frac{q}{p}\right|^2|\abarf|^2\right)\cosh(yt) 
+ \left(|\af|^2 - \left|\frac{q}{p}\right|^2|\abarf|^2\right)\cos(xt) \right. 
\nonumber \\
 & & \left. \hskip0.20in +\ 
2{\rm Re}\left(\frac{q}{p}\abarf\af{}\!^*\right)\sinh(yt) -
2{\rm Im}\left(\frac{q}{p}\abarf\af{}\!^*\right)\sin(xt)\right\} 
\label{eqn:kspp_pdf1} \\
 & & \nonumber \\
\frac{dN}{dt}(\dbar) & = &  
\frac{e^{-\Gamma t}}{2}
\left\{ \left(|\abarf|^2 + \left|\frac{p}{q}\right|^2|\af|^2\right)\cosh(yt) 
+ \left(|\abarf|^2 - \left|\frac{p}{q}\right|^2|\af|^2\right)\cos(xt) \right. 
\nonumber \\
 & & \left. \hskip0.20in +\ 
2{\rm Re}\left(\frac{p}{q}\af\abarf{}\!^*\right)\sinh(yt) -
2{\rm Im}\left(\frac{p}{q}\af\abarf{}\!^*\right)\sin(xt)\right\},
\label{eqn:kspp_pdf2}
\end{eqnarray}
where ${\cal A}^{}_f$ ($\overline{\cal A}^{}_f$)
is the decay amplitude for $D^0\ra f$ ($\dbar\ra f$) decays. 
These amplitudes must be modeled, and the parameters of the model 
must also be fitted or else taken from a previous measurement. 
For a Belle analysis of $D^0\ra K^0_S\pi^+\pi^-$ decays using
976~fb$^{-1}$ of data~\cite{belle_kspp}, the 
decay model used consisted of 14 intermediate resonances 
modeled by isobars, plus $K\pi$ and $\pi\pi$ $S$-waves.
The magnitudes and phases of the isobars, and parameters
of the $S$-waves, were obtained from a separate 
time-independent fit to the data.

An important advantage of this measurement is that the fitted PDFs
(Eqs.~\ref{eqn:kspp_pdf1} and \ref{eqn:kspp_pdf2}) depend on $x$ and 
$y$ without being ``rotated'' by an unknown strong phase difference.
The results obtained by Belle~\cite{belle_kspp} 
are listed in Table~\ref{tab:kspp_results} along with the precision 
estimated for Belle~II. This precision is obtained as follows.
The statistical errors of the Belle measurement ($\sigma^{}_{\rm stat}$)
are scaled by the square root of the ratio of luminosities. The 
systematic errors are divided into two categories: 
``reducible'' errors ($\sigma^{}_{\rm syst}$) that should decrease 
with luminosity such as errors due to background modeling as determined
from control samples; and ``irreducible'' errors ($\sigma^{}_{\rm irred}$) 
that do not decrease with more data such as uncertainty in decay 
time resolution due to detector misalignment. 
The total error is calculated as 
$\sigma^{}_{\rm Belle\ II} =  
\sqrt{(\sigma^2_{\rm stat} + \sigma^2_{\rm syst})\cdot
({\cal L}^{}_{\rm Belle}/50{\rm\ ab}^{-1}) + \sigma^2_{\rm irred}}$.
These errors are conservative, as the simple scaling used does 
not account for the improved decay time resolution of the 
Belle~II vertex detector as compared to that of Belle
(see Section~\ref{section:decay-time}).

\begin{table}[hbt]
\begin{center}
\begin{tabular}{l|cccc}  
          &  $\sigma(x)$ & $\sigma(y)$ & 
$\sigma\left|\frac{\displaystyle q}{\displaystyle p}\right|$ 
& $\sigma(\phi)$ \\
          & ($10^{-2}$)   & ($10^{-2}$)    &       &  ($^\circ$)  \\
\hline
Belle $\sigma_{\rm stat}$ & 0.19   & 0.15   & 0.155   & 10.7  \\
Belle $\sigma_{\rm syst}$ & 0.06   & 0.06   & 0.054   &  4.5  \\
Belle $\sigma_{\rm irred}$ & 0.11   & 0.04   & 0.069   & 3.8  \\
\hline
Belle II $\sigma_{\rm tot}$ & 0.11   & 0.05   & 0.073   & 4.1  \\
\hline
\end{tabular}
\end{center}
\vskip-0.10in
\caption{Precision obtained for $x$, $y$, $|q/p|$, and $\phi$ 
from a Belle analysis of $D^0\ra K^0_S\pi^+\pi^-$ decays~\cite{belle_kspp}, 
and the precision expected for Belle~II as obtained by scaling the Belle 
errors (see text).} 
\label{tab:kspp_results}
\end{table}

\vskip0.15in
\noindent {\large {\bf Acknowledgements}}
\vskip0.10in

The authors thank the CKM 2016 organizers for a well-run workshop 
and excellent hospitality. This research is supported by the 
U.S.\ Department of Energy.


\begin{thebibliography}{99}
\bibitem{charge-conjugates}
Throughout this paper, charge-conjugate modes are implicitly
included unless stated otherwise.
\end{thebibliography}
\end{document}